\documentclass[namedreferences]{solarphysics}
\usepackage{graphics,graphicx,multirow,cancel,url}
\usepackage[hyperref,optionalrh]{spr-sola-addons} 

\newcommand{\aap}{    {\it Astron. Astrophys.}}

\newcommand{\apj}{    {\it Astrophys. J.}}
\newcommand{\apjl}{   {\it Astrophys. J. Lett.}}

\newcommand{\solphys}{{\it Solar Phys.}}
 
\newcommand{\ssr}{    {\it Space Sci. Rev.}} 
\newcommand{\zap}{    {\it Zeitschrift für Astrophysik.}}

\usepackage{epsfig}          
\usepackage{graphicx,color}        
\usepackage{amssymb} 
\usepackage{color}           
\usepackage{setspace}
\usepackage{rotating}
\usepackage{float}
\usepackage{textcomp}
\usepackage{natbib}
\begin{document}

\begin{article}

\begin{opening}

\title{Statistical study of network jets observed in the solar transition region: A comparison between coronal holes and quiet sun regions}
\author{Nancy ~\surname{Narang}$^1$\sep
 Rebecca T. ~\surname{Arbacher}$^{2, 3}$ \sep
 Hui  ~\surname{Tian}$^{4}$ \sep
  Dipankar  ~\surname{Banerjee}$^{1, 5}$ \sep
  Steven R. ~\surname{Cranmer}$^6$ \sep
  Ed E.  ~\surname{DeLuca}$^2$ \sep 
  Sean ~\surname{McKillop}$^2$ }

\runningauthor{N. Narang et al.}
\runningtitle{Statistical study of network jets}
   \institute{$^{1}$ Indian Institute of Astrophysics, Koramangala 2nd Block, Bangalore 560034\\
                     email: \url{nancy@iiap.res.in}, \url{dipu@iiap.res.in}}
                     
   \institute{$^{2}$ Harvard-Smithsonian Center for Astrophysics, Garden Street, Cambridge}
   
   \institute{$^{3}$ Department of Physics, Columbia University
   }
   
   \institute{$^{4}$ School of Earth and Space Sciences, Peking University\\
   email: \url{huitian@pku.edu.cn}
   }
   
   \institute{$^{5}$ Center of Excellence in Space Sciences, IISER Kolkata 
   }
   
   \institute{$^{6}$ Department of Astrophysical and Planetary Sciences, Laboratory of Atmospheric and Space Sciences, University of Colorado 
   }


\date{Accepted: XXXX 2016, Received: \today}

\begin{abstract}
Recent IRIS observations have revealed a prevalence of intermittent small-scale jets with apparent speeds of  80 - 250 km s$^{-1}$, emanating from small-scale bright regions inside network boundaries of coronal holes. We find that these network jets appear not only in coronal holes but also in quiet-sun regions. Using IRIS 1330~{\AA} (C~{\small II}) slit-jaw images, we extract several parameters of these network jets, {\it e.g.} apparent speed, length, lifetime and increase in foot-point brightness. Using several observations, we find that some properties of the jets are very similar but others are obviously different between the quiet sun and coronal holes. For example, our study shows that the coronal-hole jets appear to be faster and longer than those in the quiet sun. This can be directly attributed to a difference in the magnetic configuration of the two regions with open magnetic field lines rooted in coronal holes and magnetic loops often present in quiet sun. We have also detected compact bright loops, likely transition region loops, mostly in quiet sun. These small loop-like regions are generally devoid of network jets. In spite of different magnetic structures in the coronal hole and quiet sun in the transition region, there appears to be no substantial difference for the increase in foot-point brightness of the jets, which suggests that the generation mechanism of these network jets is likely the same in both regions.
\end{abstract}

\end{opening}


\section{Introduction}\label{sec:intro}
The solar chromosphere and transition region (TR) act as an interface between the relatively cool photosphere ($\sim6\times10^{3}$ K) and hot corona ($\sim10^{6}$ K), and hence play a key role in the formation and acceleration of solar wind. Numerous investigations are being carried out to understand where the solar wind originates and how it is accelerated (for recent reviews see~\citet{cranmer09} and~\citet{hansteen12}). Dark regions in coronal images indicate coronal holes (CHs), which are the commonly accepted large-scale source regions of high-speed solar wind. On the other hand, the quiet sun (QS) regions (QS is a generic term for regions too bright to be coronal holes, but too dim to be regarded as magnetically \lq active\rq) are considered to be one possible large-scale source of the low-speed component of solar wind \citep{habbal01,he07,tian11}. However, identification of precise origin sites of the two different components of solar wind in the respective regions is still a challenging task as it requires high-resolution observations of the chromosphere and TR.\\

CHs are regions of low density plasma in the solar corona that have magnetic fields opening freely into interplanetary space. The existence of coronal holes was first recognized in the late 1950's, when M.~Waldemeier \citep{waldmeier56} noticed long-lived regions of negligible intensity in images made with a visible light coronagraph. In coronal images, CHs appear dark in comparison to QS because they emit less in ultraviolet and X-rays and are maintained at a lower temperature than the surrounding QS region. The different magnetic structures of CH and QS at coronal heights are responsible for their different appearance in coronal lines \citep{wiegelmann04,tian08}. CHs are dominated by open magnetic field lines expanding super-radially in the heliosphere, whereas QS regions are dominated by closed magnetic loops of different sizes.\\
 
Although CHs emit significantly less at coronal temperatures than QS regions, one can hardly distinguish between the two in most of the chromospheric and lower TR lines. Typically, lines from ions formed around 10$^{4}$ K sample the chromosphere, which is characterized by a cell-like pattern of \lq{super-granular\rq} surface flows. The super-granular network is the same in CHs and QS. As the temperature increases past 10$^{5}$ K in the thin and chaotic TR, CHs become distinguishable as areas of lower density and temperature. Also in spectroscopic observations, line parameters in the two regions differ only for TR and coronal lines. For instance,~\citet{wang13} used spectroscopic observations of the \textit{Solar Ultraviolet Measurements of Emitted Radiation} instrument (SUMER,~\citet{wilhelm95}) onboard the \textit{Solar and Heliospheric Observatory} (SOHO) spacecraft to investigate the doppler shifts and non-thermal line widths of various lines spanning the solar atmosphere from the chromosphere to the TR. It was observed that most of the TR region lines in network regions are broader and more blue-shifted in CH than in QS. On the other hand, no such distinction between CH and QS was observed for the chromospheric lines.\\

High-resolution observations from the \textit{Interface Region Imaging Spectrograph} (IRIS:~\citep{pontieu14} have revealed unprecedented levels of details in the less studied solar TR, the layer between chromosphere and corona. In a  recent work by~\citet{tian14}, detection of prevalent small-scale, high-speed jets with TR temperatures from the network structures of CHs is reported. These network jets can be significantly observed in slit-jaw images (SJIs) of Mg~{\small II} 2796~{\AA} (10$^{4}$ K, chromosphere), C~{\small II} 1330~{\AA} ($3\times10^{4}$ K, lower TR) and Si~{\small IV} 1440~{\AA} ($8\times10^{4}$ K, TR) passbands of IRIS, though best visible in C~{\small II}  1330~{\AA}. They can also be identified in the clean spectra of Si~{\small IV} lines and the most obvious signature of the network jets is the significant broadening of the line profiles. They originate from small-scale bright regions, often preceded by foot-point brightenings.~\citet{tian14} concluded that the existence of network jets are consistent with the \lq{Magnetic Furnace Model\rq} of solar wind \citep{axford92,tu05,yang13} and thus may serve as strong candidates for supply of mass and energy to the solar wind and corona.~\citet{tian14} also suggest that some of these network jets may be the on-disk counterparts and TR manifestations of the chromospheric type-II spicules (see also~\citet{pereira14,pontieu14a,rouppe15,skogsrud15}).\\

The network jets in CHs, as reported in~\citet{tian14}, have widths of $\leqslant$300 km with lifetimes of 20-80 s and lengths of 4-10 Mm. They calculated apparent speeds of the jets, which turned out be 80-250 km s$^{-1}$ in CHs. We observe that the network jets appear not only in CHs but also in QS areas. In this paper, we undertake a comparative study of various properties of these short-lived, small-scale network jets in CH and QS using IRIS SJIs in the 1330~{\AA} passband.\\
\section{Data Analysis}\label{sec:data}
\subsection{Details of Observations}
\begin{figure}[htbp]
 \vspace*{2cm}
\centering
\includegraphics[height=0.5\textheight,viewport= 65 140 600 550,keepaspectratio]{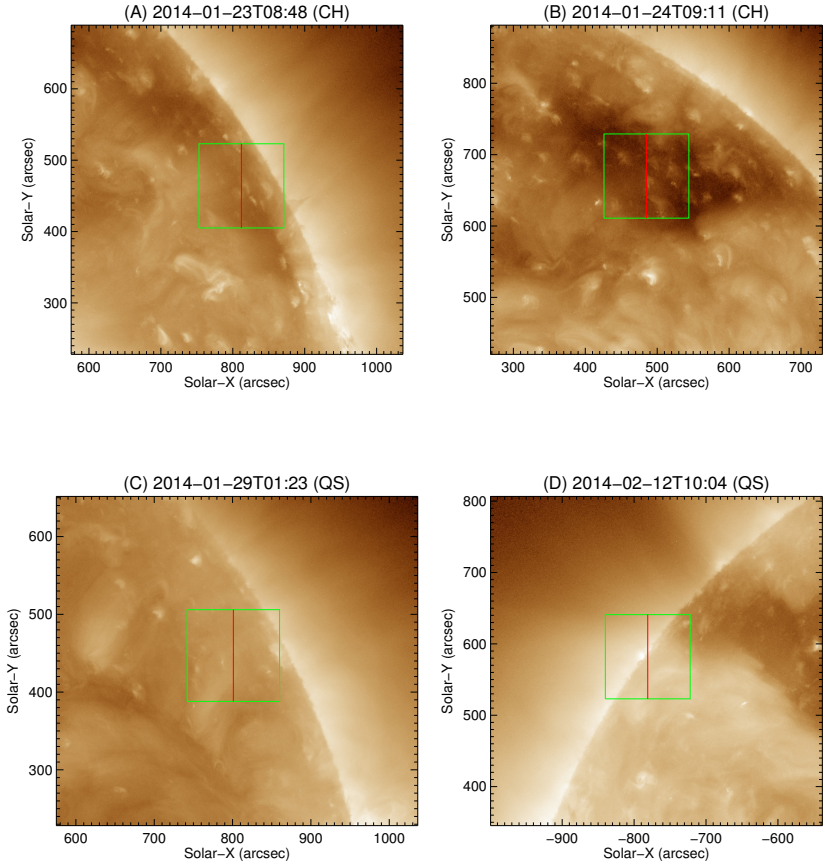}
\caption{IRIS observation regions outlined in AIA 193~{\AA} images. In each panel, the green rectangle outlines the field of view (FOV) of the IRIS SJIs and the red line shows the position of the slit. The details of the four observations shown here are provided in Table~\ref{details_obs}.}
\label{context}
\end{figure}

\begin{figure}[htbp]
 \vspace*{3cm}
 \centering
  \includegraphics[height=0.5\textheight,viewport= 50 110 650 550,keepaspectratio]{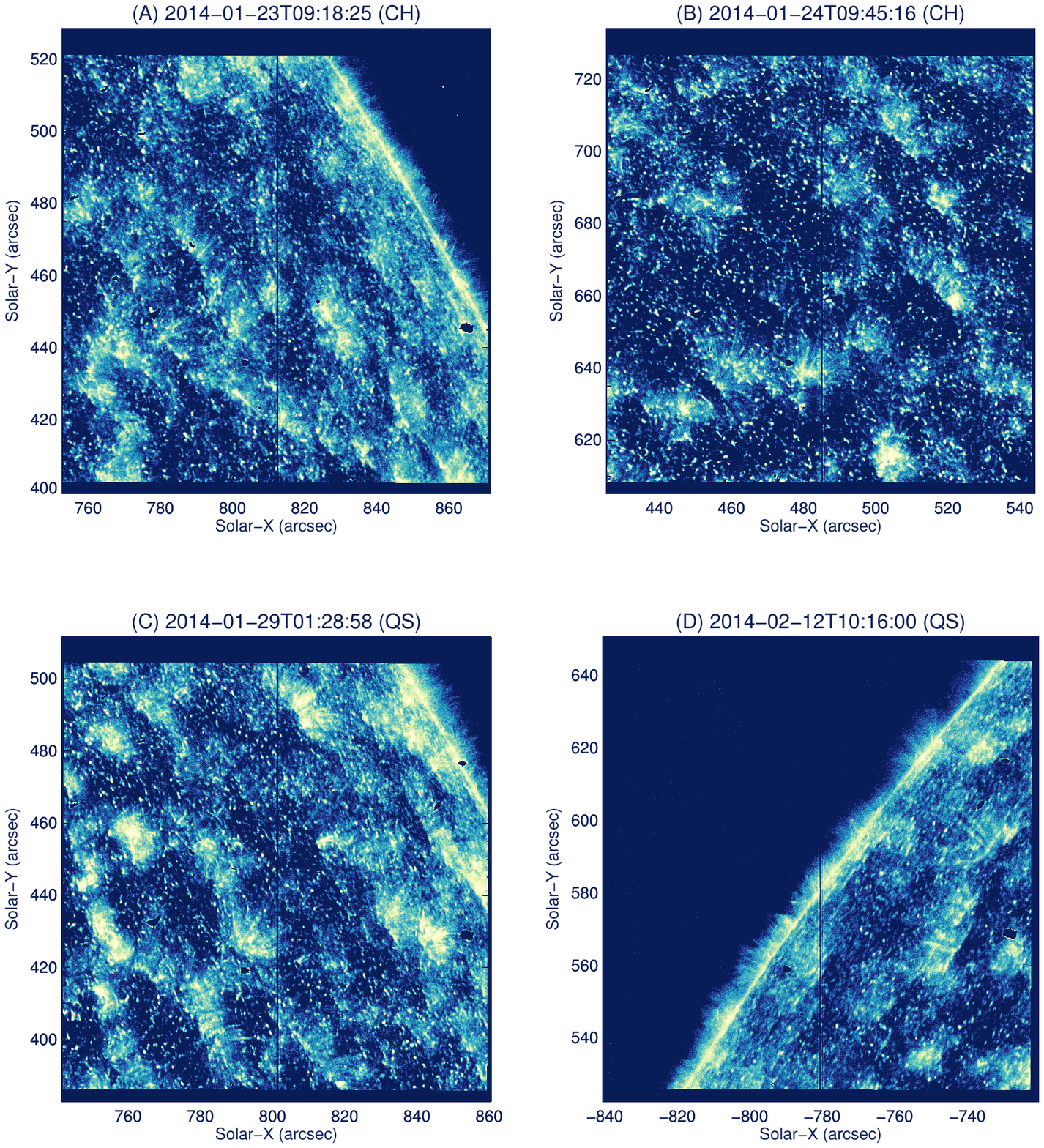}
  \caption{One of the unsharp-masked SJIs in the image sequence of each data-set used (refer to Table~\ref{details_obs} and Figure~\ref{context}). A full-FOV movie of the whole image sequence of the original and processed images of data-set (A) can be accessed in the online journal. Also see the supplementary materials of the article for the small FOV movies (zoomed) of the data-sets (B) and (C) showing the dynamics of the jets more clearly.}
\label{SJIs}
\end{figure}

Four data-sets obtained with IRIS are used in this paper: two of which are of CH and two of QS. Details of the observations are showcased in Table~\ref{details_obs}. Note that all four data-sets are high cadence sit-and-stare observations, which is essential to study the dynamics of the short-lived jets. In Figure~\ref{context}, the observed regions on the solar disk are shown as rectangles outlined in the coronal images taken in the 193~{\AA} pass-band of the \textit{Atmospheric Imaging Assembly} (AIA:~\citep{lemen12} on-board the \textit{Solar Dynamics Observatory} (SDO). The calibrated level 2 data of IRIS is used in our study. Dark current subtraction, flat field correction and geometrical correction have all been taken into account in level 2 data \citep{pontieu14}.~\citet{tian14} mentioned that the network jets are best visible in 1330~{\AA} images. Hence, in this paper we use SJIs in this passband only (see Figure~\ref{SJIs}). The 1330~{\AA} passband of IRIS samples emission from the strong C~{\small II} 1334/1335~{\AA} lines formed in the lower TR ($\sim3\times10^{4}$ K). Though, being a broad passband, it also includes UV continuum emission formed in the upper photosphere. For instance, the observed ubiquitous grain-like structures in SJIs are probably the UV emission from granules (magneto-convective cells on the solar photosphere) and acoustic or magneto-acoustic shocks (\citet{carlsson97,rutten99,sykora15}, also refer to the supplementary materials of~\inlinecite{tian14} for a detailed discussion). We explain in Section 2.3 that it is important to choose close-to-limb observations to minimize projection effects. There are not many close-to-limb observations with a large field of view and high cadence in 1330~{\AA} passband available. Moreover, the sensitivity of the IRIS far-ultravoilet SJI detector has been found to decrease significantly in the past two years. Hence the observations used  here, taken in January and February 2014, are the best for a comparative study of the very narrow network jets.\\

\begin{table}[h]
\caption{Summary of sit-and-stare observations used in this article. For all the data-sets mentioned here, field of view is $119^{\prime\prime}\times119^{\prime\prime}$, exposure time is 4 s and cadence is 10 s.}
\label{details_obs}
\begin{tabular}{cccc}
\hline
Data-Set & Observation Time &  Target &  Pointing (X,Y) \\
 \hline

A & 2014-01-23  & CH & $812^{\prime\prime},464^{\prime\prime}$\\
& 08:48 to 09:39 UT & &\\
\hline
B & 2014-01-24 & CH & $485^{\prime\prime},670^{\prime\prime}$ \\
& 09:11 to 10:06 UT & &\\ 
 \hline

C & 2014-01-29 &   QS & $801^{\prime\prime},447^{\prime\prime}$ \\
& 01:23 to 02:21 UT & & \\
 \hline

D & 2014-02-12 & QS & $-781^{\prime\prime},582^{\prime\prime}$ \\
& 10:04 to 10:46 UT & & \\

 \hline
\end{tabular}
\end{table}


\subsection{Unsharp Masking}
As mentioned earlier, due to the presence of quickly evolving grain-like structures and background network emission, the visibility of the network jets is obstructed by a considerable amount. In addition, the jets mostly appear very close to each other in space. They are also observed to recur at the same locations very often in the whole image sequences. Thus, generally it is a bit difficult to isolate individual jets in SJIs. To make the jet structures appear sharper in the images, we have applied the unsharp-masking technique to the SJIs. The technique is explained as follows: For every image in the sequence, a {$1^{\prime\prime}\times1^{\prime\prime}$} (6$\times$6 pixels) box-car smoothed version of that image is subtracted from the original image. This residual is then added back to the original image. The resultant image is referred to as unsharp masked image. This technique is also applied by~\citet{tian14} to enhance the fine features. Figure~\ref{SJIs} shows snapshots of the unsharp masked images in the four observations. Note that, due to the similar spatial and time scales, the bright grain-like features are still present in the unsharp masked images.\\

\subsection{Space--time Plots}
The technique of space--time (S--T) plots is widely used to derive the apparent speeds of moving features. The observed regions are large in size and hence we can identify many network jets in each of the observations shown in Figure~\ref{SJIs}. It should be noted that all four observed regions are close to the limb and thus the line of sight component of the jet velocities should be small. Hence, the value of apparent speeds derived from S--T plots must be close to the real velocities of the jets.\\

We have visually identified 31 jets in data-set (A), 36 in (B) (making a total of 67 jets in CHs) and 52 jets in data-set (C) and 8 in (D) (making a total of 60 jets in QS regions); with relatively strong emission in the 1330~{\AA} image sequences. The selection of jets is made in such a way that every such jet is well isolated from others in space and time. A lot of care has been taken so that the selected jets are less affected by the bright grain-like structures and thus show clear signatures in the S--T plots. It is important to emphasise here that we are able to measure only the relatively strong jets which is only small fraction (roughly~$\sim$20-30\%) of the network jets present in the data. Also note that the data-set D has only eight jets that can be reliably traced because the duration of this data-set is less (see Table~\ref{details_obs}) and, most important, there is so much off-limb in the field of view in comparison to other data-sets (see Figure~\ref{SJIs} (D)).\\

\begin{figure}
\centering
\includegraphics[width=1.2\textwidth,viewport=50 460 650 700, keepaspectratio]{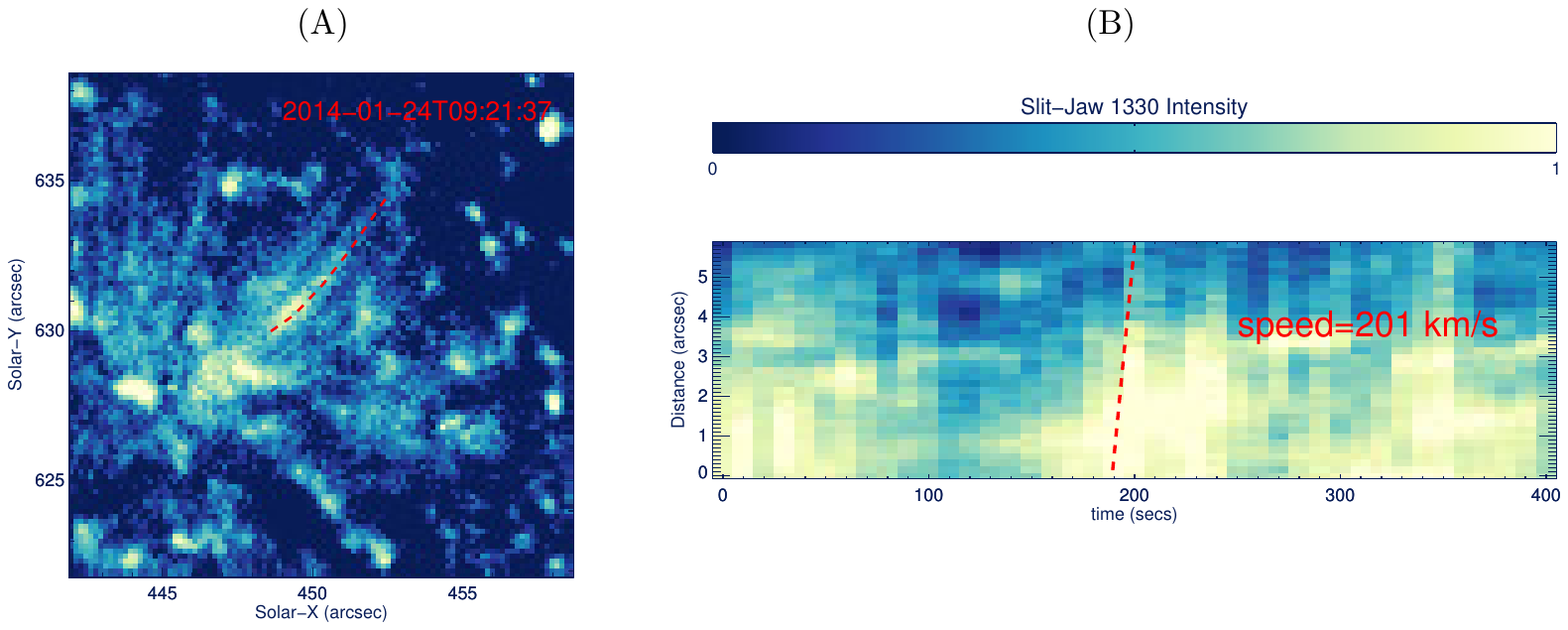}

\includegraphics[width=1.2\textwidth,viewport=50 460 650 700, keepaspectratio]{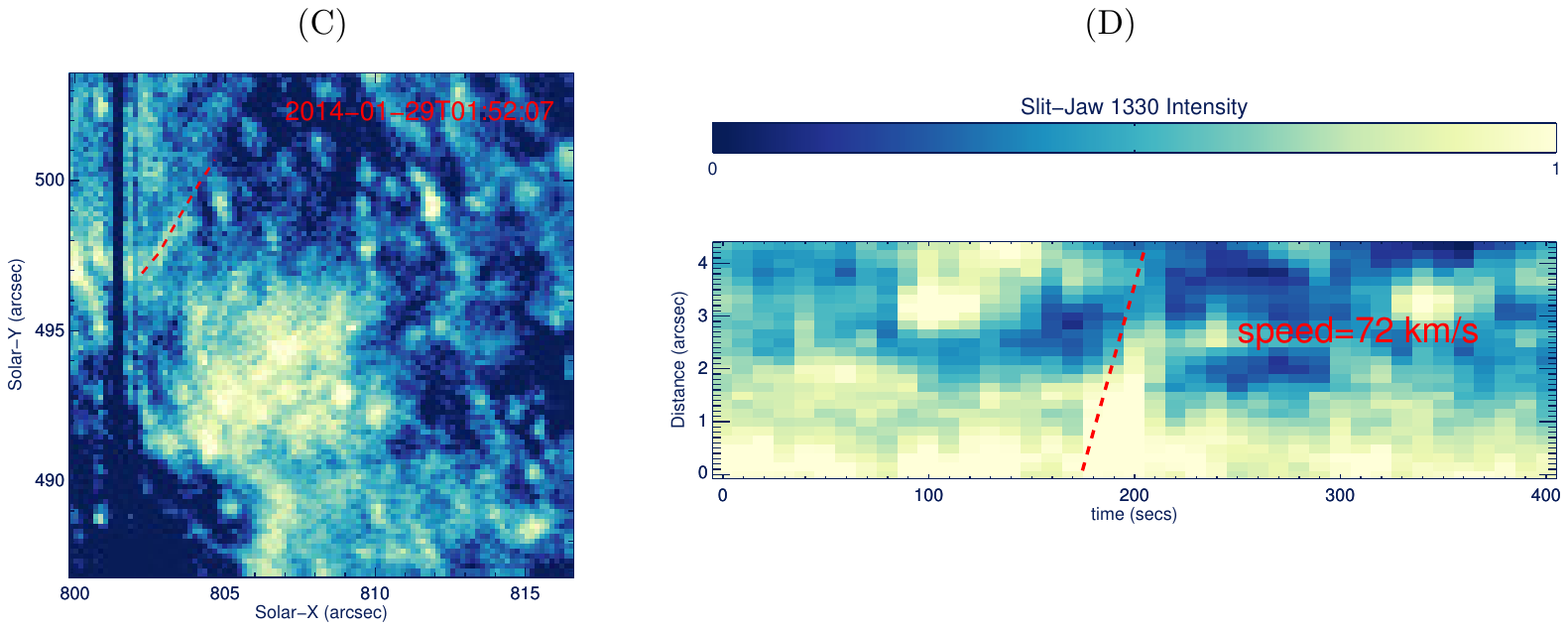}
\caption{(A) A small FOV of one of the unsharp-masked images of dataset-B (refer to Figure~\ref{SJIs}). The dashed line indicates the path of a jet. (B) S--T map for the jet marked in (A). (C) Same as (A) but for dataset-C. (D) S--T map for the jet marked in (C).}
\label{cq}
\end{figure}

For each jet identified, we first draw a line (curved or straight) along the direction of propagation of the jet (see example in Figure~\ref{cq} (A) and (C)). The intensity along this line is plotted and then stacked with time (Figure~\ref{cq} (B) and (D)). The lifetime and maximum length of the jet can be obtained directly from the S--T map. The measurements done here are solely based on visual inspection. Due to weak emission of the jets and complications by the network grains, it is very difficult to use any automatic method. The minimum lifetime observed is 20 s, although it may be possible that many jets have lifetimes shorter than 20 s as the observations are limited by a cadence of 10 s. The apparent speed can be calculated as the slope of the inclined strip in the S--T plot. For example, the apparent speed of the jet marked in Figure~\ref{cq} (A) and (B) is calculated to be 201 km s$^{-1}$ and that in Figure~\ref{cq} (C) and (D) is calculated to be 72 km s$^{-1}$. Total of four data-sets are analyzed in the present work, independently by the first two authors. Results obtained by them are generally consistent. It is important to note that the uncertainty in the measurement of speed is dictated by the spatial resolution and time cadence of the data. The spatial resolution of IRIS SJIs is $\sim$250 km and cadence of all the data-sets used is 10 s, hence the minimum uncertainty in the calculation of speeds of the jets turns out to be $\sim$25 km s$^{-1}$. \\

\subsection{Foot-point Brightness}
It is already mentioned that the network jets are mostly preceded by foot-point brightenings. As it is important to explore the impact of local heating on the dynamical properties of the jets, we calculated the increase in foot-point brightness for every jet. The foot-point of jet is defined as the location of origin of the jet. We choose a 2$\times$2 pixel area around the foot-point of the jet. The intensity within this area is determined at the instant, closest to appearance of the jet, when the brightness at the foot-point is maximum, say {\it brightness}. The intensity at the same location is determined in two more frames in the image sequence, one before and one after the frame of maximum intensity at the foot-point. These two frames are selected such that there is no enhanced brightening at the location of the foot-point of the jet as compared to surrounding area. The average of the intensities at the location of the foot-point of these two frames is calculated, say {\it average}. This {\it average} is subtracted from the {\it brightness} (as termed above). The resultant is divided by the {\it average} and multiplied by 100 to get result in terms of a percentage. Note that by term intensity used above throughout, we mean the total data-counts in the 2$\times$2 pixel area obtained from original SJIs.\\


\section{Results and Discussion}\label{sec:result}
\begin{figure}
 \vspace*{5cm}
\includegraphics[width=1.0\textwidth,viewport= 40 175 600 400,keepaspectratio]{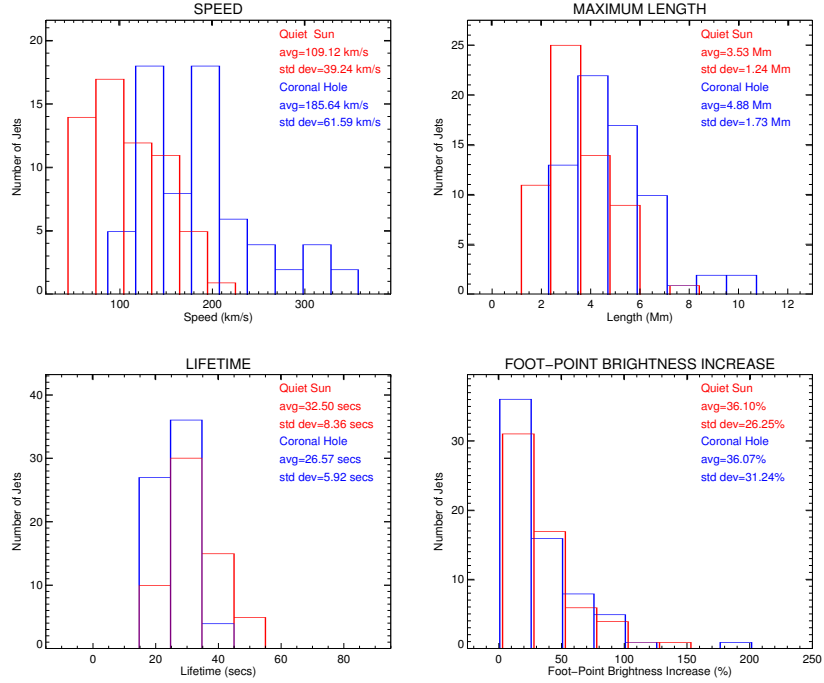}
\caption{Distribution of different parameters of network jets showing comparison between CH and QS regions. The total sample number for CH distributions (indicated by blue colour) is 67 jets and that for QS distributions (indicated by red colour) is 60 jets.}
\label{hist}
\end{figure}

We have analyzed different properties of the network jets (mainly apparent speed, length, lifetime and increase in foot-point brightness) to study their dynamics and for comparison between CH and QS jets. The comparison clearly shows that the average values of apparent speed and length of jets in CHs are significantly greater than those in QS regions as showcased in Figure~\ref{hist}. The QS results are marked in red and CH results are represented by blue histograms. Though, there does not exist any such demarcation for cases of lifetime and foot-point brightness increase. The comparison of calculated average values (with standard deviation) of the above mentioned properties is summarized in Table~\ref{result}.\\

\begin{table}[h]
\caption{Average properties of network jets. The errors mentioned are standard deviations of the respective distribution.}
\label{result}

\begin{tabular}{ccc}
\hline
 & {\bf Coronal Hole} & {\bf Quiet Sun} \\ \hline
{\bf Speed (km s$^{-1}$)} & $186\pm62$ & $109\pm39$ \\
{\bf Length (Mm)} & $4.9\pm1.7$ & $3.5\pm1.2$ \\ 
{\bf Lifetime (s)} & $27\pm6$ & $33\pm8$ \\ 
{\bf F-P Brightness Increase (\%)} & $36\pm31$ & $36\pm26$ \\ \hline
\end{tabular}
\end{table}

As CH jets appear to be faster and longer than QS jets, this may be a consequence of different magnetic configurations in the CH and QS areas. In CHs, open and expanding magnetic flux tubes at network boundaries must be assisting the small-scale network jets to propagate up to larger extents with higher speeds and, hence the jets are accelerated more efficiently in CHs than in QS regions. This result is consistent with a recent numerical simulations on chromospheric jets by~\citet{iijima15}. They have pointed out that these jets are projected farther outward with higher speeds when overlying coronal gas pressure is lower (similar to that in coronal hole) and shorter when the coronal gas pressure is higher (similar to that in quiet sun) which  agrees with our observations.  We should point out that in their simulation jets are  generated by chromospheric shocks and the amplification of chromospheric shock wave will be different in CH and QS. Our observed greater apparent speeds of these jets in CHs as compared to the QS also allows to explore the suggestions made by~\citet{tian14} and~\citet{rouppe15} that some network jets are likely to be TR manifestations of rapid blueward excursions (RBEs). It has already been claimed ({\it e.g.}~\citealp{rouppe09,sekse12,pereira14}) that the RBEs observed in profiles of different chromospheric lines are on-disk counter parts of solar spicules. In addition, it is also reported that the doppler velocity of RBEs increases along their length in CHs when observed in the 8542~{\AA} (Ca~{\small II}) spectral line using the \textit{CRisp Imaging SpectroPolarimeter instrument} (CRISP,~\citet{scharmer08}) at the \textit{Swedish 1-m Solar Telescope} (SST). However, no such trend was observed for QS RBEs (see~\citealp{sekse13} for detailed discussion). This correspondence between the increasing trend of doppler velocity of RBEs (observed in the chromosphere) and higher speeds of the network jets (observed in the TR) in CHs provide more evidence that the RBEs, or on-disk spicules may be the signatures of lower-temperature and less-accelerated parts or phases of the network jets.\\

\begin{figure}[h]
\includegraphics[width=1.05\textwidth,viewport= 50 360 600 700,keepaspectratio]{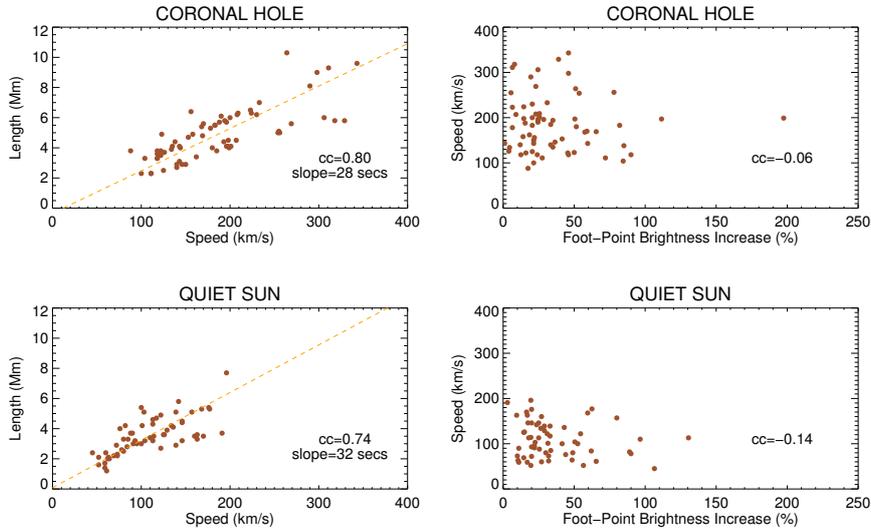}
\caption{Scatter plots for different parameters of network jets. The correlation coefficients (cc) are indicated.}
\label{plots}
\end{figure}


The distribution and average value of foot-point brightness increase is almost the same in CH and QS. The similar magnitude probably suggests the same generation mechanism of the network jets in both the regions.  In addition, we have studied the inter-relation between the jet properties in respective regions in Figure~\ref{plots}. We find that the apparent speed of the network jets is independent of increase in foot-point brightness, although it is very much dependent on length of the jets. As indicated in Figure~\ref{plots}, speed and length of the jets are highly correlated in both CH and QS regions with a correlation coefficient of 0.80 for CH and 0.74 for QS jets. On the other hand, brightness increase at foot-points appears to be independent of any of the dynamical properties of the jets. This again reflects that the basic mechanism ({\it e.g.} magnetic reconnection) responsible for generation of the network jets is of very similar nature in CH and QS. However, as the jets propagate in different ambient magnetic environments in the two regions, most of the dynamical properties get considerably affected.\\

The recurrence of these high speed network jets from the same location suggests that the oscillatory reconnection might be a possible mechanism for generation of the jets. It has already been demonstrated by~\citet{murray09} and~\citet{mclaughlin12} that such a reconnection can trigger quasi-periodic upflows. It is also reported that the Lorentz-force driven models are able to produce, heat and accelerate spicules \citep{sykora11,goodman14}. The recent numerical simulation by~\citet{goodman14} even claims that the Lorentz-force driven jets can have speeds as high as 66-397 km s$^{-1}$, similar to speeds found in the present study. However, the pressure driven jets ({\it e.g.}~\citet{sykora11,judge12}) are reported to gain speeds $\sim60$ km s$^{-1}$ only. Thus, based on the above mentioned numerical simulation results along with our observations, it can be concluded that the evolution of magnetic field at small scales has a key role in the generation and acceleration of high-speed jets in the chromosphere and TR. It is important to note that the optically thin Si~{\small IV} lines do not have enough S/N in the 4s exposure data, so the spectral data can not be used in such observations. Without detailed analysis of the spectroscopic data, we can not exclude the possibility that some of the apparent motions may be reflection of the ionization front or even shock waves and are likely not real mass flows (see~\citealp{tian14} for a detailed discussion).\\

\begin{figure}
\includegraphics[width=0.5\textwidth,keepaspectratio,angle=90]{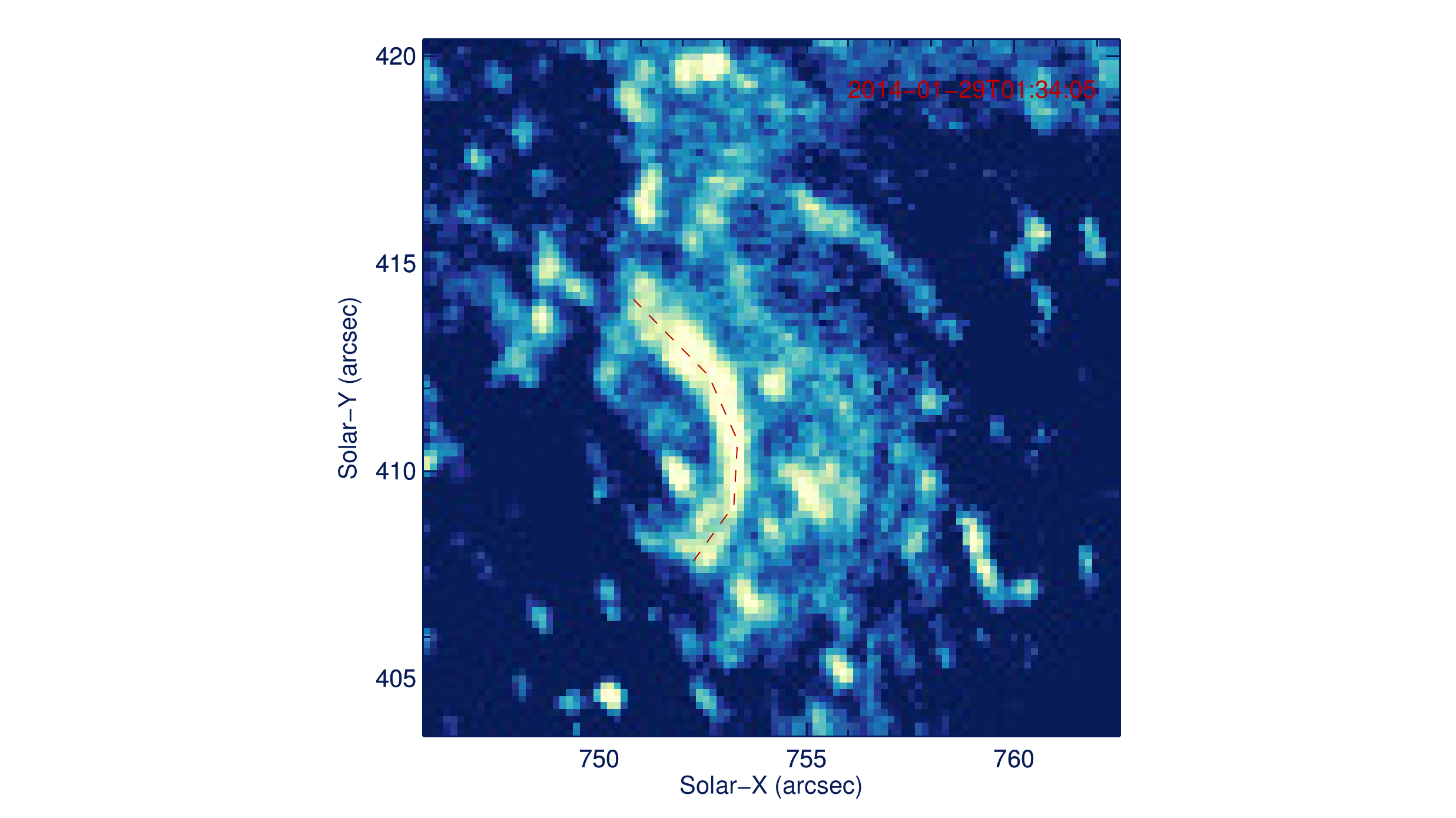}
\caption{One example of compact bright loop observed in QS. The red curve marks the loop-like feature.}
\label{loop}
\end{figure}

We have already indicated that the mechanism generating of the network jets should be similar in the CH and QS. If the generation mechanism is magnetic reconnection in the chromosphere (\textit{e.g.}~\citealp{shibata07,yurchyshyn13,deng15,ni15}), it would indicate that there is no significant difference of magnetic structures in the chromospheric layers of CH and QS. In CH, it is probably small chromospheric loops that reconnect with open flux in the network. In case of QS, small chromospheric loops reconnect with the legs of the coronal loops. Higher up in the TR and corona, there are probably not many loops in the CH. While in QS, there are still a lot of TR and coronal loops present. Our observations provide direct imaging evidence to validate this proposed idea as we have found some small compact loop-like bright features to be present in QS (likely the TR loops reported by~\citet{hansteen14}). An example of such a loop detected in our observations in QS is showcased in Figure~\ref{loop}. These bright loop-like regions have typical extents of $\sim5^{\prime\prime}$ and are generally devoid of the network jets. In CH, no such features can be observed, suggesting that, at the layers of TR in a CH, there are basically only open field lines and almost no loops. This result is also consistent with the findings of~\citet{wiegelmann04,tian08} that loops reside only at very low layers in a CH. The difference in magnetic morphology in higher layers is likely responsible for the different propagation of the network jets in the two regions, leading to higher speed and longer distance in CHs. The fact that the observed QS compact loops are generally devoid of the network jets actually suggests that the network jets occurred below the height of these TR loops, which means that the network jets are likely produced in the chromosphere.\\ 
In addition to explanations involving magnetic reconnection, there are some other possible models that could account for the observed properties
of the jets.~\citet{hollweg82} proposed that magnetohydrodynamic waves could exert a time-averaged upward nonthermal pressure and thus ``levitate'' the cool chromosphere. This was suggested as a formation mechanism for large, Type I solar spicules (see also~\citealp{depontieu99,kudoh99,matsumoto10}).
~\citet{cranmer15} recently extended this idea to smaller-scale turbulent Alfv\'{e}nic motions in CH network regions, and found that nonlinear mode conversion into chromospheric shocks may provide intermittent upflows of the right order of magnitude. Such a model produces vertical excursions in the height of the transition region of order 1 to 6 Mm over timescales of 20 to 60 s, which gives rise to apparent jet-like velocities of 50 to 200 km s$^{-1}$. The wave/turbulence model for jet formation is observationally distinguishable from a reconnection-based model, in that the former does not require magnetic field of both polarities to be present at the jet footpoint location.\\

The prevalence of the network jets and that the jets reach to higher layers ($\sim5$ Mm) in both the regions (CH and QS), implies that they may play an important role in supplying mass and energy to the corona and solar wind. At this stage, one must reconcile the implications on the origin of the solar wind \citep{tian14} due to presence of these intermittent small-scale network jets in the TR. However, IRIS observations are unable to detect heating signatures of these network jets at coronal temperatures as IRIS is designed to study mainly the chromosphere and TR and over its spectral coverage, there is only one line present, Fe~{\small XII} 1349~{\AA}, which is formed at normal coronal temperatures but it is usually very weak or absent. The observations show only that the jets reach at least $\sim10^{5}$ K, but presently we cannot account if they can be heated to coronal temperatures. In the near future, we will try to track these jets to coronal structures. One way, for instance, would be to investigate the possible connection between these jets and the hot intermittent upflows along plume-like structures \citep{mcintosh10,tian11,pucci14,pant15} and the blue-shift patches of lower coronal lines \citep{tian081,tian09,fu14}. Moreover, one must also try to address the question of dissipation of energy by the network jets in the corona, {\it e.g.}, whether they trigger compressional waves or not \citep{gupta12,uritsky13,jiao15,pant15}. This aspect can be studied in more details in order to further explore the difference in mechanisms between CH and QS for generating network jets and for processing the supplied mass and energy by these jets to the corona and solar wind. In this context, new instrumentations with similar spatial and spectral resolution as that of IRIS, along with wider wavelength coverage, that can provide co-spatial and co-temporal observations of the TR and corona, are also desirable. For instance, the \textit{Extreme Ultravoilet Imager} (EUI) and \textit{Spectral Imaging of the Coronal Environment} (SPICE) instruments onboard the \textit{Solar Orbiter} spacecraft \citep{muller13}, to be launched in October 2018, can provide more insight into the heating process of the network jets. This will serve as a great opportunity for a better understanding of the relationship between the TR network jets and coronal structures and outflows.\\


\section{Summary}\label{sec:summary}
Our IRIS observations reveal the presence of network jets in QS as well as in CH. We have conducted some comparative analysis between CH and QS regions based on the properties of these jets. It must be noted that the results obtained are limited by the number of data-sets used for the study. As mentioned earlier, we have used only two observations in CH and two in QS regions. Hence, presently we are not sure if the results will change by using more observations. Our results from the current study are summarized as follows:

\begin{enumerate}
  \item CH jets appear to be faster and longer than those in QS regions. This is most likely a consequence of different magnetic configurations of the two regions with open magnetic field lines dominant in CH and magnetic loops often present in QS. This proposed idea is well supported by our observations which clearly show some compact bright loops to be present in QS but generally absent in CH at TR heights.
  
  \item Recently, RBEs are reported to show an increasing trend in the doppler velocity along their length in CH. The higher apparent speed of network jets in CH can provide evidence to support the proposed idea that TR network jets are the accelerated phase of RBEs in coronal holes. 
  
  \item The similar distribution and average value of increase in foot-point brightness indicates towards the same generation mechanism of these jets in both regions. Moreover, we find a good correlation between apparent speed and length of the jets. On the other hand, the foot-brightness increase seems to be independent of any of the dynamical properties of the jets.
  
  \item  As these jets reach up to higher layers (length of $\sim5$ Mm), they can serve as reasonable candidates for supplying mass and energy to the corona and solar wind. However, it is important to note that IRIS observations are unable to detect the signatures of these jets beyond $\sim10^5$ K.
  
\end{enumerate}
  
\section*{Acknowledgments}
We would like to thank the anonymous referee for his/her constructive comments which has enabled us to improve the manuscript. Authors are thankful to the IRIS team for making the data  publicly available. IRIS is a NASA Small Explorer mission developed and operated by the Lockheed Martin Solar and Astrophysics Laboratory (LMSAL), with mission operations executed at the NASA Ames Research Center and major contributions to downlink communications funded by the Norwegian Space Center (Norway) through a European Space Agency PRODEX contract. This work was supported by contract 8100002705 from LMSAL to SAO and the NSF REU solar physics program at SAO, grant number AGS-1263241. N.~N. would like to thank Human Resource Development Group (HRDG) of Council of Scientific and Industrial Research (CSIR), India for awarding Junior Research Fellowship (JRF). H.~T. is supported by the Recruitment Program of Global Experts of China and NSFC under grant 41574166. We would like to thank Haruhisa Iijima for valuable suggestions.\\

{\bf \centering SUPPLEMENTARY MATERIAL\\}

The following movies are available in the electronic version of the journal (refer to Figure~\ref{SJIs} in main text).\\
{\bf Movie(A).} Full FOV image Sequence of data-set(A).\\
{\bf Movie(B).} A small FOV (zoomed) image sequence of data-set(B).\\
{\bf Movie(C).} Same as Movie(B) but for data-set(C).



\end{article} 

\end{document}